\documentclass[journal,11pt,onecolumn]{IEEEtran}
\usepackage{amsmath}
\usepackage{amsfonts}
\usepackage{amsthm}
\usepackage{dsfont}
\usepackage{epsfig}
\usepackage{upgreek}
\usepackage{graphicx}

\usepackage{setspace}

\theoremstyle{definition}

\newcommand{\RGA}{\mbox{${\mathcal RGA}$}}
\newcommand{\MPRGA}{\mbox{${\mathcal MP}$-${\mathcal RGA}$}}
\newcommand{\UCRGA}{\mbox{${\mathcal UC}$-${\mathcal RGA}$}}

\newcommand{\Imat}{\mbox{${\bf I}$}}

\newcommand{\Amn}{\mbox{${\bf A}$}}

\newcommand{\Bm}{\mbox{${\bf B}$}}

\newcommand{\Tpose}[1]{\mbox{${#1}^{\mbox{\scriptsize{T}}}$}}
\newcommand{\Perm}{\mbox{${\bf P}$}}

\newcommand{\Qm}{\mbox{${\bf Q}$}}

\newcommand{\Vm}{\mbox{${\bf V}$}}
\newcommand{\Vmi}{\mbox{${\bf V}^*$}}
\newcommand{\Xm}{\mbox{${\bf X}$}}

\newcommand{\Ym}{\mbox{${\bf Y}$}}

\newcommand{\Mm}{\mbox{${\bf M}$}}
\newcommand{\Dm}{\mbox{${\bf D}$}}
\newcommand{\Dmi}{\mbox{${\bf D}^{\mbox{\tiny -1}}$}}

\newcommand{\Em}{\mbox{${\bf E}$}}
\newcommand{\Emi}{\mbox{${\bf E}^{\mbox{\tiny -1}}$}}

\newcommand{\Tm}{\mbox{${\bf T}$}}

\newcommand{\Gm}{\mbox{${\bf G}$}}

\newcommand{\Gmi}{\mbox{${\bf G}^{\mbox{\tiny -1}}$}}
\newcommand{\Gmp}{\mbox{${\bf G}^{\mbox{\tiny -P}}$}}

\newcommand{\ginv}[1]{{#1}^{\mbox{\tiny -U}}}
\newcommand{\pinv}[1]{{#1}^{\mbox{\tiny -P}}}

\newcommand{\inv}[1]{{#1}^{\mbox{\tiny -1}}}

\newcommand{\Um}{\mbox{${\bf U}$}}
\newcommand{\Umi}{\mbox{${\bf U}^*$}}

\newcommand{\Atinv}{\mbox{${\bf A}^{\overset{\sim}{\!\mbox{\tiny -1}}}$}}

\newcommand{\Apinv}{\mbox{${\bf A}^{\!\mbox{\tiny -P}}$}}
\newcommand{\Aginv}{\mbox{${\bf A}^{\!\mbox{\tiny -U}}$}}

\newcommand{\tinv}[1]{\mbox{${#1}^{\overset{\sim}{\mbox{\tiny -1}}}$}}

\begin{document}

\title{{On the Relative Gain Array (RGA) with\\
         Singular and Rectangular Matrices} }       
\author{
%{\large\bf Jeffrey Uhlmann}\\
%University of Missouri-Columbia}
\IEEEauthorblockN{{Jeffrey Uhlmann}}\\
\IEEEauthorblockA{\small University of Missouri-Columbia\\
201 Naka Hall, Columbia, MO 65211\\
573.884.2129, uhlmannj@missouri.edu}}
\date{}          
\maketitle
\thispagestyle{empty}

%For singlespaced version
%\vspace{-16pt}
%For doublespaced version
%\vspace{-0.5in}

\begin{abstract}
This paper identifies a significant deficiency in the 
literature on the application of the Relative Gain Array (RGA)
formalism in the case of singular matrices. Specifically, it is
shown that the conventional use of the Moore-Penrose pseudoinverse
is inappropriate because it
fails to preserve critical properties that can be assumed in
the nonsingular case. It is then shown that such properties
can be rigorously preserved using an alternative generalized matrix
inverse.\\ 

\begin{footnotesize}
\vspace{-8pt}
\end{footnotesize}
\end{abstract}

\section{Introduction}

The relative gain array (RGA) provides a jointly-conditioned relative
measure of input-output interactions
among a multi-input multi-output (MIMO) system\!~\cite{bristol}. 
The RGA is a function of a matrix, which typically is interpreted
as a plant/gain matrix $\Gm$ for a set of input parameters that control 
a set of output parameters, and is defined for nonsingular $\Gm$ as:
\begin{equation}
   \RGA(\Gm) ~\doteq~ \Gm \circ \Tpose{(\Gmi)}
\end{equation}
where $\circ$ represents the elementwise Hadamard matrix product.

The RGA has been and continues to be used in a wide variety of
important practical control applications. This is due in part to its
convenient use and interpretation and in part to its established
mathematical properties. The RGA is also applied widely in a form 
that is generalized for applications with singular $\Gm$ and is presumed 
to maintain at least some of the rigor and properties that hold in the 
nonsingular case\!~\cite{skog,changyu}. In this paper
it is shown that this presumption is unjustified and therefore
should undermine confidence in existing safety-critical systems
in which it is applied. We then propose an alternative 
generalization that provably maintains a key property which 
does not hold for the convential formulation presented in the 
literature.

\section{Properties of the RGA}

When the gain matrix $\Gm$ is nonsingular the RGA
has several important properties~\!\cite{rgajs,HJ2}, 
which include for 
diagonal matrices $\Dm$ and $\Em$ and permutation
matrices $\Perm$ and $\Qm$:
\begin{eqnarray}
   \RGA(\Perm\Gm\Qm) & = & \Perm\cdot\RGA(\Gm)\cdot\Qm\\
   \RGA(\Dm\Gm\Em) & = & \RGA(\Gm).
\end{eqnarray}
The first property simply says that a permutation, i.e., reordering, of the 
elements of the input and output vectors leads to a conformant reordering
of the rows and columns of the resulting RGA matrix. The second
property says that the relative gain values are invariant to the units
applied to elements of the input and output vectors. In other words,
the relative gain is independent of the choice of units chosen for 
state variables, e.g., celsius versus farenheit for temperature variables 
or radians versus degrees for angle variables. 

Both of the above properties represent intuitively natural sanity
checks because there should certainly be grave concerns if the
integrity of a system were impacted by the choice of how the
parameters/variables are ordered or the choice of units used
for those parameters/variables. The value of the RGA is that it
provides a unit-invariant measure of the sensitivity of each output
variable to each input parameter\footnote{The RGA actually provides
more than just a measure of sensitivity, e.g., the ratio of open-loop 
gain to closed-loop gain is a very useful interpretation for controller 
pairing.}. For example, if the output variables
are redefined from imperial units to metric units the resulting controls 
obtained from the RGA will produce {\em identical} system behavior. 

Of course the definition of the RGA as $\Gm \circ \Tpose{(\Gmi)}$
requires $\Gm$ to be nonsingular to ensure the existence of its inverse.
The most obvious approach for generalizing the definition to include
the case of singular $\Gm$ is to replace the matrix inverse $\Gmi$ with a
generalized inverse $\tinv{\Gm}$ such that, ideally, the result:
\begin{equation}
   \RGA(\Gm) ~\doteq~ \Gm \circ (\tinv{\Gm}\Tpose{)}
\end{equation}
preserves the key properties satisfied for nonsingular $\Gm$.
Universally in the RGA literature the generalized inverse is taken
to be the Moore-Penrose (MP) pseudoinverse\!~\cite{skog,changyu,nrrga}, 
which is denoted here as $\Gmp$. This choice seems to derive from a 
widely-held misperception that the MP inverse is somehow uniquely 
suited to be the default whenever a generalized inverse is 
required\!~\cite{uhlmann0}. The fact is that the MP inverse
is only applicable under certain sets of assumptions, and in the
case of the RGA its use will not generally preserve the key
property of unit invariance. This will be demonstrated in the 
following section, where it will also be shown that there
exists an alternative generalized inverse that does preserve 
the unit-invariance property.

\section{Generalized Matrix Inverses}

Although the generalized matrix inverse of a singular matrix $\Amn$ as 
$\Atinv$ cannot satisfy all properties of a true inverse, it is typically
expected to satisfy the following algebraic identities:
\begin{eqnarray}
      \Amn\Atinv\Amn & = & \Amn,\\
      \Atinv\Amn\Atinv\!\! & = & \Atinv.
\end{eqnarray}
Beyond these properties, the appropriate generalized 
inverse of choice for a particular application must be 
determined by the conditions it is expected to preserve. 
In the case of the MP inverse, it satisfies the following:
\begin{equation}
     \pinv{(\Um\Amn\Vm)} ~ = ~ \Vmi\Apinv\Umi \label{udist}
\end{equation}
where $\Um$ and $\Vm$ are arbitrary rotation matrices (or, more generally,
arbitrary orthonormal/unitary matrices) such that $\Um\Umi=\Imat$, where
$\Umi$ is the conjugate-transpose of $\Um$. This consistency with
respect to rigid rotations is necessary when Euclidean distances must
be preserved by all state space transformations. This preservation of
the Euclidean norm is also what makes it appropriate for determining
the $n\times m$ matrix $\Tm$ representing the minimum mean-squared
error (MMSE) optimal 
linear transformation, $\Tm\Xm\approx\Ym$, where the $n$-dimensional
columns of rectangular matrix $\Xm$ represent a set of $p$ input vectors 
and the $m$-dimensional columns of $\Ym$ represent a set of $p$ output
vectors. 

The MP inverse is intimately connected to ``least-squares'' optimization, and both
are commonly used in scientific and engineering applications without
regard to whether the MMSE criterion is appropriate. In the case of
vectors with elements that represent state variables, e.g., that include a mix of
quantities such as temperature, mass, velocity, etc., the minimization
of squared deviations, or equivalently the preserving of Euclidean
distances between vectors, has no physical meaning. The problem
for least-squares is, of course, that the magnitude of squared 
deviations is strongly dependent on the choice of units~\cite{zhang}. In the case
of a range-bearing estimate $(r,\theta)$, for example, the choice to 
represent $r$ in centimeters rather than meters, or even kilometers, 
has many orders of magnitude implications when squared
deviations are minimized. The bearing angle $\theta$, by contrast,
is likely to be constrained between -$\pi$ and $\pi$, or between
$0$ and $2\pi$, so depending on the units chosen for $r$ the
deviations in angle may be relatively so small as to be
negligible from a least-squares perspective.

The unit-consistent (UC) generalized inverse\!~\cite{uhlmann}, 
denoted as $\Aginv$, satisfies a different consistency condition:
\begin{eqnarray}
      \ginv{(\Dm\Amn\Em)}  & = & \Emi\Aginv\Dmi
\end{eqnarray}
where $\Dm$ and $\Em$ are arbitrary nonsingular diagonal
matrices. Thus it satisfies
\begin{equation}
   \Em\cdot\ginv{(\Dm\Amn\Em)}\cdot\Dm ~=~ \Aginv
\end{equation}
while the MP inverse does not. As suggested in~\cite{uhlsc} this is 
precisely the condition required to ensure that a
generalized definition of the RGA:
\begin{equation}
   \UCRGA(\Gm) ~\doteq~ \Gm \circ (\ginv{\Gm}\Tpose{)}
\end{equation}
preserves the unit-invariance property. Specifically, it is straightforward
to show that the UC-RGA is invariant with respect to the scaling of
its argument $\Gm$ by nonsingular diagonal matrices $\Dm$ and $\Em$:
\begin{eqnarray}
   \UCRGA(\Dm\Gm\Em) & = & 
          (\Dm\Gm\Em) \circ (\ginv{(\Dm\Gm\Em)}\Tpose{)} \\
    ~ & = & (\Dm\Gm\Em) \circ (\Emi\ginv{\Gm}\Dmi\Tpose{)} \label{uc}\\
    ~ & = & (\Dm\Gm\Em) \circ (\Dmi(\ginv{\Gm}\Tpose{)}\Emi) \\
    ~ & = & (\Dm\Dmi)\left(\Gm \circ (\ginv{\Gm}\Tpose{)}\right)(\Em\Emi) \\
    ~ & = & \Gm \circ (\ginv{\Gm}\Tpose{)} \\
    ~ & = & \UCRGA(\Gm) 
\end{eqnarray}
where the step of Eq.\!~(\ref{uc}) exploits the unit-consistency property
of the UC inverse, which is not satisfied by the MP inverse. 

\section{Properties of the UC-RGA}

The UC-RGA and the RGA are equivalent when $\Gm$ is nonsingular because
$\ginv{\Gm} =\inv{\Gm}$. Therefore, it is the case of singular $\Gm$ that
remains to be examined. In the $2\times 2$ case singular $\Gm$ is either 
the zero matrix or has rank $1$, i.e., the system
is constrained to one dimension and therefore interactions among the two
input-output pairings are scalar-equivalent. The rank-1 case is degenerate
but illustrative because the UC-RGA is always 
\begin{equation}
   \UCRGA(\Gm) ~=~ \begin{bmatrix}0.25 & 0.25\\0.25 & 0.25\end{bmatrix}, 
\end{equation}
where the row and column sums are the same in analogy to the
general case for the RGA of a nonsingular matrix. This is not generally
the case for singular $\Gm$, as can be inferred from 
\begin{equation}
   \Gm ~=~ \begin{bmatrix}a & b\\0 & 0\end{bmatrix}, 
\end{equation}
for which the second row of $\RGA(\Gm)$ must also be zero
and therefore cannot have the same sum as the columns
(assuming $a$ and $b$ are nonzero). What can be 
said in general is that the sum of the
elements of $\RGA(\Gm)$ will equal the rank of $\Gm$, and the
fact that the sum of every row and column equals $1$ when 
$\Gm$ is nonsingular is due to the fact that its rows and 
columns are linearly independent.

A revealing example of the difference between the
UC-RGA and the MP-RGA is the case of 
\begin{equation}
   \Gm ~=~ \begin{bmatrix}
                        1 & 1 & 1\\
                        1 & 1 & 1\\
                        1 & 1 & 1
                  \end{bmatrix}, 
\end{equation}
for which the interactions among inputs and outputs
clearly should be the same. In this case both the UC-RGA
and MP-RGA can be verified to yield the same result
\begin{equation}
    \frac{1}{9}\cdot
    \begin{bmatrix}
          1 & 1 & 1\\
          1 & 1 & 1\\
          1 & 1 & 1
    \end{bmatrix}
\end{equation}
where, as expected, all of the interaction values are 
the same (and sum to $1$, the rank of $\Gm$).
However, if the first row and column are scaled
by $2$:
\begin{equation}
   \Gm ~=~ \begin{bmatrix}
                        4 & 2 & 2\\
                        2 & 1 & 1\\
                        2 & 1 & 1
                  \end{bmatrix}, 
\end{equation}
the UC-RGA result is unaffected whereas
the MP-RGA gives
\begin{equation}
   \frac{1}{9}\cdot
                 \begin{bmatrix}
                        4 & 1 & 1\\
                        1 & 1/4 & 1/4\\
                        1 & 1/4 & 1/4
                  \end{bmatrix}, 
\end{equation}
which demonstrates clear unit-scale dependency. The
more general case of rectangular $\Gm$ can be sanity-checked
using an arbitrary nonsingular $3\times 3$ matrix 
\begin{equation}
   \Amn ~=~ \begin{bmatrix}
                        7 & 4 & 8\\
                        7 & 2 & 5\\
                        3 & 8 & 8
                  \end{bmatrix} 
\end{equation}
and
\begin{equation}
   \Bm ~=~ \begin{bmatrix}
                        21 & 16 & 16\\
                        21 & 8 & 10\\
                        9 & 32 & 16
                  \end{bmatrix}, 
\end{equation}
which are equivalent up to diagonal scaling of their columns
and therefore have the same RGA, which can be verified to 
be\footnote{Numeric values in this example have been rounded
to two decimal places. It should be noted that the values in the 
example matrices of this section are chosen to be integers
purely for simplicity in replicating results using
code provided in the appendix. The matrices are not intended to
represent or be interpreted as meaningful plant matrices, though
realistic examples can be generated and examined using code from
the appendix.}:
\begin{equation}
   \RGA(\Amn)=\RGA(\Bm) = \begin{bmatrix}
                        \text{-}2.47 & \text{-}2.41 & ~5.88\\
                        ~3.29 & ~0.94 & \text{-}3.24\\
                        ~0.18 & ~2.47 & \text{-}1.65
                  \end{bmatrix}. 
\end{equation}
Thus, one should expect that UC-RGA applied to the block-rectangular
matrix $\Mm=[\Amn~\Bm]$ will produce a result in 
which its two $3\times 3$ rectangular blocks are identical, and it does
indeed satisfy this expectation:\\

\begin{equation}
\UCRGA(\Mm) ~=~
   \frac{1}{2}\cdot
      \begin{bmatrix}
         \text{-}2.47 & \text{-}2.41 & ~5.88 & \text{-}2.47 & \text{-}2.41 & ~5.88\\
         ~3.29 & ~0.94 & \text{-}3.24 & ~3.29 & ~0.94 & \text{-}3.24\\
         ~0.18 & ~2.47 & \text{-}1.65 & ~0.18 & ~2.47 & \text{-}1.65
       \end{bmatrix}.
\end{equation}

\vspace{6pt}
\noindent The MP-RGA result, by contrast, fails this sanity test:\\

\begin{equation}
\MPRGA(\Mm) ~=~
   \frac{1}{2}\cdot
      \begin{bmatrix}
         \text{-}4.47 & \text{-}4.54 & ~9.41 & \text{-}0.49 & \text{-}0.28 & ~2.35\\
         ~5.93 & ~1.77 & \text{-}5.18 & ~0.66 & ~0.11 & \text{-}1.29\\
         ~0.32 & ~4.65 & \text{-}2.64 & ~0.04 & ~0.29 & \text{-}0.66
       \end{bmatrix}
\end{equation}
as its first $3\times 3$ block is completely different from 
the second $3\times 3$ block despite the fact they are
derived from matrices that are identical up to a scaling of columns
and should both equal the matrix of Equation~(25).

\section{Discussion}

This paper has provided evidence that the Moore-Penrose (MP) pseudoinverse
is inappropriate for use to generalize the relative gain array (RGA) for
applications involving singular matrices. Given the fact that it has been
used in this way for decades, it is natural to ask how such an error 
could have gone unnoticed. It turns out that questions have been 
raised in the literature. In~\cite{nrrga}, for example, the authors 
express a caveat regarding use of the MP inverse for RGA 
``control in the least-square sense'' and that its implications 
``should be investigated.'' 

It is also reasonable to speculate that self-selection has discouraged
publication of reports of failed applications of the MP-RGA and that
published reports of successful applications may have omitted
discussion of engineering efforts applied to mitigate
unexplained performance issues. For example, units may have 
been changed judiciously so that (implicitly) input-output variables
receive comparable weight under the least-squares 
criterion\footnote{Various forms of ``normalization''
are used heuristically in many engineering domains to 
address unit-dependency issues for which rigorous solutions exist.
Such heuristics are often regarded as ordinary aspects of 
implementation engineering and consequently may not 
be reported.}. 

It is hoped that the unit-consistent (UC) solution described
in this paper will promote renewed interest in applications
of the generalized RGA and, more importantly, provide
greater rigor and reliability to resulting systems. 

\appendices

\section{Code for Proposed RGA Solution}

\noindent The following is an Octave/Matlab implementation
of the generalized RGA (adapted from an implementation of
the UC-inverse from\!~\cite{uhlmann}). The function takes
a matrix argument and returns the generalized RGA.
\begin{verbatim}
function R = ucrga(A)
    tol = 1e-15;    
    [m, n] = size(A);
    L = zeros(m, n);    M = ones(m, n);   
    S = sign(A);   AA = abs(A);
    idx = find(AA > 0.0);  L(idx) = log(AA(idx));
    idx = setdiff(1 : numel(AA), idx);
    L(idx) = 0;    M(idx) = 0;   
    r = sum(M, 2);   c = sum(M, 1);   
    u = zeros(m, 1); v = zeros(1, n);
    dx = 2*tol;  
    while (dx > tol)
        idx = c > 0;
        p = sum(L(:, idx), 1) ./ c(idx);
        L(:, idx) = L(:, idx) - repmat(p, m, 1) .* M(:, idx);
        v(idx) = v(idx) - p;  dx = mean(abs(p));
        idx = r > 0;
        p = sum(L(idx, :), 2) ./ r(idx);
        L(idx, :) = L(idx, :) - repmat(p, 1, n) .* M(idx, :);
        u(idx) = u(idx) - p;  dx = dx + mean(abs(p));
    end    
    dl = exp(u);   dr = exp(v);
    S = S.* exp(L);
    R = A .* transpose(pinv(S) .* (dl * dr)');
end
\end{verbatim}


\begin{thebibliography}{7}

\bibitem{nrrga}
K.H.\  Baeka, T.F.\ Edgar, K.\ Song, G.\ Choi, H.K.\ Cho, and C.\ Han,
``An effective procedure for sensor variable selection and utilization in
plasma etching for semiconductor manufacturing,''
{\em Computers and Chemical Engineering}, 61, 20\text{-}29, 2013.

\bibitem{ben}
Ben-Israel, Adi; Greville, Thomas N.E., ``Generalized inverses: Theory and Applications," 
{\em New York: Springer}, 2003.

\bibitem{bristol}
E. Bristol, ``On a new measure of interaction for multivariable process
control,'' {\em IEEE Transactions on Automatic Control}, vol.\ 11, no.\ 1, pp.\
133-134, 1966.

\bibitem{changyu}
J. Chang and C. Yu, ``The relative gain for non-square multivariable
systems,'' {\em Chemical Engineering Science}, vol. 45, no.\ 5, pp.\ 1309-1323,
1990.

\bibitem{hanson}
Robert Hanson, ``Matrices Whose Inverse Contain Only Integers,''
{\em The Two-Year College of Mathematics Journal}, Vol\ 13, No.\ 1, 1982.

\bibitem{HJ2}
R.A.\ Horn and C.R.\ Johnson, {\em Topics in Matrix Analysis}, 
Cambridge University Press, 2007.

\bibitem{rgajs}
C.R.\ Johnson and H.M.\ Shapiro, ``Mathematical Aspects of the Relative Gain 
Array $(A \circ A^{-T})$'', 
{\em SIAM. J. on Algebraic and Discrete Methods}, 7(4), 627-644, 1985.

\bibitem{skog}
S.\ Skogestad and I.\ Postlethwaite, {\em Multivariable Feedback Control:
Analysis and Design}, 2nd ed., New York: Wiley 2005.

\bibitem{uhlsc}
Jeffrey Uhlmann, ``A Rank-Preserving Generalized Matrix Inverse for Consistency with Respect to Similarity,''  
{\em IEEE Control Systems Letters}, ISSN: 2475-1456, 2018.

\bibitem{uhlmann0}
Jeffrey Uhlmann,  ``Unit Consistency, Generalized Inverses, and
Effective System Design Methods,'' arXiv:1604.08476v2 [cs.NA] 11 Jul 2017 (2015).

\bibitem{uhlmann}
Jeffrey Uhlmann, ``A Generalized Matrix Inverse that is Consistent with Respect to Diagonal Transformations,''
{\em SIAM Journal on Matrix Analysis} (SIMAX), Vol.\ 39:3, 2018.

\bibitem{zhang}
Bo Zhang and Jeffrey Uhlmann, ``Applying a New Generalized Matrix Inverse
For Stable Control of Robotic Systems,'' arXiv:1806.01776v1 [eess.SP] 7 May, 2018.


\end{thebibliography}
\end{document}